\documentclass[aps,prd,notitlepage,nofootinbib,superscriptaddress,eqsecnum,twocolumn,floatfix]{revtex4-1}

\usepackage[utf8]{inputenc}
\usepackage{amsmath}
\usepackage{amssymb}
\usepackage{amsfonts}
\usepackage{newtxtext,newtxmath}
\usepackage{bm}
\usepackage{graphicx}

\usepackage[usenames,dvipsnames]{xcolor}
\usepackage{color}
\usepackage[colorlinks=true,linkcolor=blue,urlcolor=blue,citecolor=blue]{hyperref}
\usepackage{slashed}
\usepackage[english]{babel}
\usepackage{dcolumn}
\usepackage{blindtext}
\usepackage{epsfig}
\usepackage{pifont}
\usepackage{dsfont,mathrsfs}
\usepackage{cancel}
\usepackage{bigints}
\usepackage{accents}
\usepackage{soul}
\usepackage{multirow}
\usepackage{ulem}
\usepackage{subfigure} 
\usepackage{simpler-wick}
\usepackage{float}

\begin{document}

\title{Speed of sound peak in two-color dense QCD:\\ Confronting effective models with lattice data}

\author{Arthur E. B. Pasqualotto}\email{ arthur.pasqualotto@acad.ufsm.br}
\affiliation{Departamento de F\'{\i}sica, Universidade Federal de Santa
  Maria,  97105-900 Santa Maria, RS, Brazil}

\author{ Dyana C. Duarte}\email{ dyana.duarte@ufsm.br}
\affiliation{Departamento de F\'{\i}sica, Universidade Federal de Santa
  Maria, 97105-900 Santa Maria, RS, Brazil}
  
\author{ Ricardo L. S. Farias}\email{ ricardo.farias@ufsm.br}
\affiliation{Departamento de F\'{\i}sica, Universidade Federal de Santa
  Maria, 97105-900 Santa Maria, RS, Brazil}
\affiliation{Center for Nuclear Research, Department of Physics, Kent State University, Kent, Ohio 44242 USA}  
  
\author{ Rudnei O. Ramos} \email{ rudnei@uerj.br}
\affiliation{Departamento de F\'{\i}sica Te\'orica, Universidade do Estado do Rio de Janeiro, 20550-013 Rio de Janeiro, RJ, Brazil}
\begin{abstract}
Lattice simulations of two-color, two-flavor quantum chromodynamics
(QCD) at finite quark chemical potential have revealed a distinctive
peak structure in the sound velocity. Although chiral perturbation theory (ChPT)
and the Nambu-Jona-Lasinio (NJL) model have been employed to explain
this phenomenon, neither approach has fully captured the observed
behavior.  To address this discrepancy, we have extended the NJL
framework by incorporating the medium separation scheme (MSS). This
approach isolates medium contributions from divergent integrals,
allowing for a more accurate treatment of finite-density effects. Our
results indicate a clear increase in the diquark gap ($\Delta$) with increasing chemical potential, consistent with what is also seen in perturbative QCD predictions at high densities. {}Furthermore, the MSS-modified NJL model
successfully reproduces the observed peak in the sound velocity.

\end{abstract}

\maketitle

\section{Introduction}

Quantum chromodynamics (QCD), the theory governing strong
interactions, presents significant theoretical challenges due to its nonperturbative nature at relevant energy scales~\cite{Gross:2022hyw}. Constructing a comprehensive phase diagram for QCD requires understanding both the confinement of quarks and gluons at low energies and the asymptotic freedom at high energies, as well as chiral symmetry breaking~\cite{Fukushima:2025ujk}. Although
QCD lattice simulations have successfully mapped the phase transition between these distinct phases at zero density and finite temperature~\cite{Aoki:2006we}, the equation of state (EoS) for low-temperature QCD at finite density remains a formidable obstacle due to the sign problem~\cite{Aarts:2015tyj,Nagata:2021ugx}.

The equation of state (EoS) plays a crucial role in understanding the transition from hadronic matter to a deconfined phase of quarks and gluons~\cite{Aoki:2006we,Bhattacharya:2014ara}. The EoS provides a comprehensive description of equilibrium QCD, relating thermodynamic quantities such as pressure, energy
density, and entropy. Its phenomenological importance extends beyond QCD, encompassing a wide range of systems from heavy-ion physics to astrophysics and cosmology~\cite{Teaney:2001av,Kolb:2003dz,Boyanovsky:2006bf,Lattimer:2000nx,Tews:2018kmu,Annala:2019puf}.

The precise behavior of the sound velocity in QCD matter remains an open question. Various theoretical approaches have been successfully applied to investigate the properties of both low- and high-density nuclear matter. In the low-density regime, nucleons are the dominant
degrees of freedom, and chiral perturbation theory (ChPT) provides a suitable framework in this domain~\cite{Tews:2012fj}. At extremely high baryonic densities, asymptotic freedom enables perturbative QCD (pQCD) to establish an upper bound for the speed of
sound, approaching the conformal limit $c_s^2 = 1/3$ from
below~\cite{Kurkela:2009gj}. The region of low temperature and finite baryonic density is particularly important for understanding the phase structure of QCD, including phenomena such as the deconfinement phase transition and the quarkyonic phase~\cite{McLerran:2007qj,McLerran:2018hbz,Duarte:2020xsp,Duarte:2020kvi,Fujimoto:2025sxx}.

The square sound speed is defined as the adiabatic derivative of the pressure density $P$ with respect to the energy density $\varepsilon$,
\begin{align}
c_s^2 = \frac{\partial P}{\partial \varepsilon}
\label{cs2dPdE}.
\end{align}
As a derivative of the EoS, the behavior of the sound velocity in this phenomenological density regime is crucial to understanding the internal structure of neutron stars~\cite{Kojo:2020krb}. The peak in sound velocity has been linked to recent observational data from multimessenger astronomy, particularly with respect to highly massive neutron stars~\cite{Tan:2021ahl}.

Monte-Carlo first-principles calculations, such as lattice QCD, have been successfully employed to describe QCD matter at zero baryonic chemical potential and finite temperature $(\mu = 0, \ T \neq 0)$. At finite isospin chemical potential, $\mu_I \neq 0$, lattice QCD
simulations have already demonstrated a peak in the speed of sound as a function of isospin density, $n_I$~\cite{Braun:2022olp,Brandt:2022hwy,Brandt:2017oyy,Wilhelm:2019fvp,Abbott:2024vhj,Abbott:2023coj}.  However, the low-temperature and finite-baryon density regime remains inaccessible to lattice simulations due to the well-known sign problem~\cite{Nagata:2021ugx}.  Analogously to the finite isospin case, the two-color QCD (QC$_2$D) is also free of the sign problem~\cite{Kogut:2001na,Buividovich:2008wf,Braguta:2023yhd,Hands:2006ve,Ilgenfritz:2013ara,Langfeld:2013xbf,Ilgenfritz:2012fw,Braguta:2016cpw,Braguta:2015zta,Braguta:2014gea}, allowing accurate predictions of various thermodynamic quantities as functions of the baryon chemical potential $\mu$, as recently demonstrated in Ref.~\cite{Iida:2024irv,Begun:2022bxj}.

Effective field theory and low-energy effective models~\cite{Adhikari:2018kzh,Kawaguchi:2023olk}, Dyson-Schwinger equations~\cite{Contant:2019lwf}, perturbative QCD~\cite{Gorda:2014vga} and renormalization group approaches~\cite{Fejos:2025nvd,Fejos:2025oxi,Dupuis:2020fhh,Khan:2015puu,Hattori:2019zig} have been widely used to investigate the qualitative behavior of QCD matter. 
Among the various effective models considered in the literature, the Nambu–Jona-Lasinio (NJL) model~\cite{Nambu:1961tp,Nambu:1961fr,Buballa:2003qv,Klevansky:1992qe,Andersen:2012jf,Kojo:2021hqh} and the linear sigma model (LSM), or the quark meson model (QMM)~\cite{Suenaga:2023xwa,Suenaga:2022uqn,Kawaguchi:2023olk,Strodthoff:2013cua,Strodthoff:2011tz} are among the frameworks used most extensively. 
The role of collective excitations in dense QC$_2$D has been systematically investigated in Refs.~\cite{Suenaga:2019jjv,Kojo:2021knn,Suenaga:2021bjz}, where the mesonic and baryonic modes were analyzed in the presence of chiral and diquark condensates. These works clarified how spectral functions and in-medium correlations encode essential nonperturbative dynamics, offering guidance for effective model constructions at high density. The LSM, in particular, has been applied to the case of QC$_2$D, providing further support for the presence of a peak structure in the finite-density regime~\cite{Kawaguchi:2024iaw,Suenaga:2025sln}.

In recent years, a series of works have demonstrated the emergence of dual symmetries between chiral and isospin imbalances, both in QC$_2$D and QC$_3$D, and their correspondence with effective models~\cite{Khunjua:2020xws,Khunjua:2021oxf,Khunjua:2024kdc}. These studies have also explored the role of chiral chemical potentials, and in~\cite{Klimenko:2025aie}, the authors explored the conditions under which charged pion condensation may occur in quark matter. More recently, duality relations in dense isotopically and chirally asymmetric QCD have been analyzed in detail, further reinforcing the usefulness of symmetry-based approaches for understanding the phase structure of strongly interacting matter~\cite{Klimenko:2024qzq}.

In this study, we use a two-color effective NJL-type model to investigate the behavior of the sound velocity. As a nonrenormalizable effective model, the cutoff regularization of divergent momentum integrals can lead to a mixing of vacuum and medium effects. 
The medium separation scheme (MSS), introduced in Ref.~\cite{Farias:2005cr} and subsequently applied in several contexts~\cite{Lopes:2025rvn,Azeredo:2024sqc,Lopes:2021tro,Avancini:2019ego,Duarte:2018kfd,Farias:2016let,Farias:2006cs,Lopes:2025rvn,daSilva:2025koa}, provides a systematic procedure to disentangle medium-dependent contributions from divergent integrals. By construction, the vacuum divergences are isolated and coincide with those of the model in the absence of a medium, being therefore unaffected by medium effects. The key principle underlying the MSS is that medium contributions are finite and can be integrated over the full momentum space, whereas the divergent vacuum terms must be separated prior to the implementation of the usual ultraviolet regularization.
We apply the MSS to the NJL model with diquark interactions as an effective description of QC$_2$D. 
By consistently separating vacuum and medium effects, this framework allows for an accurate description of the speed of sound. 
Within the MSS-modified NJL model, we successfully reproduce the characteristic peak observed in the sound speed, in agreement with lattice results.
In addition, we find that the diquark gap $\Delta$ increases monotonically with the chemical potential, consistently approaching the behavior predicted by perturbative QCD at high densities~\cite{Fukushima:2024gmp}.

The remainder of this paper is organized as follows. In Sec.~\ref{sec2}, we introduce the model considered in this work. The regularization scheme is introduced and discussed in Sec.~\ref{regularization}. The main numerical results are presented in Sec.~\ref{sec3}, while Sec.~\ref{conclusions} is devoted to the conclusions and final remarks. 
An Appendix containing further technical details of the calculation is also included.
 
%%%%%%%%%%%%%%%%%%%%%%%%%%%%%%%%%%%%%%%%%%%%%%%%%%
\section{The two-color NJL model}
\label{sec2}

In this work, we consider a two-color and two-flavor model including scalar, pseudoscalar, and color pairing interactions. The NJL Lagrangian density is given by
\begin{eqnarray}
\mathcal{L} &=& \bar \psi (i\gamma^\mu\partial_\mu-m)\psi+
G_s[(\bar\psi\psi)^2 +(\bar\psi i \gamma_5 \vec\tau
  \psi)^2]  \label{NJL_diq}%\\ &
  \nonumber\\ & &+ G_d(\bar \psi i \gamma_5 \tau_2 t_2 C
\bar\psi^T)(\psi^T C i \gamma_5 \tau_2 t_ 2 \psi),
\end{eqnarray}
where $m$ is the current quark mass, $\tau_i$ and $t_i$ are the Pauli matrices in flavor and color spaces, respectively, and the scalar and diquark coupling constants are connected by a Fierz transformation in the color space as $G_s = G_d = G$ in the two-color case~\cite{Ratti:2004ra}. In the QC$_2$D the diquarks may be understood as colorless baryons, the diquark condensation breaks the $U_B$(1) symmetry rather than the color $SU_c$(2) symmetry, while the confinement is less important~\cite{Sun:2007fc}.

The nonrenormalizability of the NJL model implies the existence of a cutoff parameter $\Lambda$, which, along with $m$ and $G$, must be fixed by an appropriate criterion. 
The most common procedure in the three-color case is to fix the parameters such as to reproduce the empirical values of the pion mass $m_{\pi}$, the pion decay constant $f_{\pi}$, and the quark condensate $\langle\bar{q}q\rangle_0$ in vacuum. To obtain the parameter values shown in Table~\ref{tab1} we considered the standard system of equations relating the parameters of the model~\cite{Buballa:2003qv,Klevansky:1992qe}
\begin{eqnarray}
    \frac{m}{M} &=& 4 G N_c N_f m_\pi^2 J(m_\pi^2) ,
    \label{eq1:system_param}\\
    f_\pi^2&=& 2 N_c N_f M^2 J(0),
     \label{eq2:system_param}\\
    \langle \bar qq\rangle_0&=& -\frac{M-m}{4G},
    \label{eq3:system_param}
\end{eqnarray}
where $J(q^2)$, with $q^2=0$ and $q^2=m_\pi^2$, is defined as
\begin{eqnarray}
    J(q^2) = \int \frac{d^4p}{(2\pi)^4}\frac{1}{(p^2-M^2)[(p+q)^2-M^2]}.
\end{eqnarray}
In solving the system of equations~(\ref{eq1:system_param})--(\ref{eq3:system_param}), we have adjusted the values of $f_\pi$ and $\langle \bar qq\rangle$
such as to reproduce the value of $m_\pi$ considered. Then the system also gives the values of the remaining parameters $G$, $m$ and $\Lambda$ along with $m_\pi$.

In this work, we employ two different parametrizations. The first corresponds to the physical pion mass, and is consistent with previous QC$_2$D analysis, e.g., Ref.~\cite{Ratti:2004ra}, while the second employs a heavier pion mass, for a better comparison with lattice QCD results from Ref.~\cite{Iida:2024irv}, since the use of a larger pion mass is a common strategy in LQCD to improve numerical convergence.
 
{}For the physical pion we follow the $N_c$-scaling of physical quantities procedure, adopted in~\cite{Brauner:2009gu,Duarte:2015ppa}, i.e., given the fact that $f_{\pi}\propto N_c^{1/2}$ and $\langle\bar{q}q\rangle_0\propto N_c$, we rescale their values by the factors $\sqrt{2/3}$ and $2/3$ respectively ($m_{\pi}$ is independent of $N_c$)\footnote{The set of empirical values we are using to rescale the physical quantities are $m_{\pi} =$ 140 MeV, $f_{\pi} = 92.4$ MeV and $-\langle\bar{q}q\rangle_0^{1/3} = 250$ MeV.}. 
In order to reliably compare the NJL model predictions with lattice results, we also parametrize the model to reproduce the pion mass used in Ref.~\cite{Iida:2024irv}.

In Table~\ref{tab1} we show the scaled values of empirical quantities \textcolor{blue}and both sets of parameters used in this work.

\begin{table*}[htb]
\caption{\label{tab:params}Parameter values used in this work. The first line corresponds to the parameters used for the physical mass parametrization, while the second line corresponds to the heavy pion mass.}
\begin{ruledtabular}
\begin{tabular}{ccccccc}
Alias & $G$ & $m$ & $\Lambda$ & $f_\pi$ & $m_\pi$ & $\langle \bar q q \rangle^{1/3}_0$ \\
\hline
Physical pion & 7.25 GeV$^{-2}$ & 5.40 MeV & 0.656 GeV & 75.45 MeV & 140 MeV & $-218$ MeV \\ Heavy pion& 2.05 GeV$^{-2}$ & 79.0 MeV & 1.262 GeV & 160.0 MeV & 738 MeV & $-447$ MeV \\
\end{tabular}
\end{ruledtabular}
\label{tab1}
\end{table*}

The mean field thermodynamic potential at finite temperature $T$ and quark chemical potential $\mu$ is given by~\cite{Sun:2007fc} 
\begin{eqnarray}
\Omega &=& \frac{(M-m)^2 + \Delta^2}{4G} - N_c N_f\sum_{s = \pm 1}\int\frac{d^3p}{(2\pi^3)}E_{\Delta}^s\nonumber\\
& & -2N_c N_f T\sum_{s = \pm 1}\int\frac{d^3p}{(2\pi)^3}\ln{\left[1 + e^{-{E_{\Delta}^s/T}}\right]},
\label{OmegaT}
\end{eqnarray}
where the modified dispersion relation is 
\begin{equation}
E_{\Delta}^s = \sqrt{\left(\sqrt{p^2 + M^2} + s\mu \right)^2 + \Delta^2}\,.
\end{equation}
In the limit $T\to 0$, which is the regime of interest in this work, the last term in the thermodynamic potential~(\ref{OmegaT}) vanishes, and we have
\begin{eqnarray}
\Omega_{T = 0} &=& \Omega_0 + \frac{(M-m)^2 + \Delta^2}{4G}\nonumber\\ & & -N_c N_f\sum_{s = \pm 1}\int\frac{d^3p}{(2\pi^3)}E_{\Delta}^s,
\label{OmegaT0}
\end{eqnarray}
where 
\begin{equation}
\Omega_0 = -\frac{(M_0-m)^2}{4G}
+ 2N_c N_f\int\frac{d^3p}{(2\pi)^3}\sqrt{p^2 + M_0^2},
\label{Omega0}
\end{equation}
is a constant added to the thermodynamic potential to produce a vanishing pressure in the vacuum, while $M_0$ is the value of the effective vacuum quark mass, which is determined at $T = \mu = 0$. The pressure density $P$, quark number density $n$, and the energy density $\varepsilon$ are defined as 
\begin{eqnarray}
P & = & - \Omega\,,\\
n & = & -\frac{\partial\Omega}{\partial\mu} = \frac{\partial P}{\partial\mu}\,,\\
\varepsilon & = & -P + \mu n\,.
\end{eqnarray}

In the next section, we discuss the different regularization procedures used in this work when evaluating the ultraviolet (UV) momentum integral in the thermodynamic potential. 

%%%%%%%%%%%%%%%%%%%%%%%%%%%%%%%%%%%%%%%%%%%%%%%%%%
\section{Regularization procedure}
\label{regularization}

Most studies employing the NJL model utilize a three-dimensional sharp cutoff $\Lambda$ to regularize the divergent momentum integrals, such as the one appearing in Eq.~\eqref{OmegaT0}. In this work, we will refer to this method as the traditional regularization scheme (TRS). Due to the nonrenormalizable nature of the model, $\Lambda$ serves as a scale for numerical calculations and must be determined alongside the current quark mass $m$ and the coupling constant $G$, as mentioned before. Although this is the most common procedure, it involves the regularization of integrals with implicit or explicit dependencies on medium effects. It may lead to inaccurate predictions regarding the evolution of the order parameters with the temperature or chemical potentials, as well as to unexpected behavior for some thermodynamic quantities. Renormalization group (RG) approaches have also been developed for low-energy effective models~\cite{Braun:2018svj} and recently used to mitigate cutoff illnesses of different effective models in the investigation of dense quark matter~\cite{Andersen:2024qus,Gholami:2024diy}. In this work, we will make use of the MSS regularization procedure~\cite{Farias:2005cr,Lopes:2025rvn,Azeredo:2024sqc,Lopes:2021tro,Avancini:2019ego,Duarte:2018kfd,Farias:2016let,Farias:2006cs} discussed in the introduction.  

%%%%%%%%%%%%%%%%%%%%%%%%%%%%%%%%%%%%%%%%%%%%%%%%%%%
\subsection{Gap equations for the diquark and chiral condensates}
\label{gaps}

The gap equations are obtained by minimizing the thermodynamic potential~\eqref{OmegaT0} with respect to $\Delta$ and $M$:
\begin{eqnarray}
\frac{\partial\Omega}{\partial\Delta}\Bigr|_{\Delta}
=0\;\;\;\;\text{ and } \;\;\;\; \frac{\partial\Omega}{\partial M}\Bigr|_{M}
=0.
\end{eqnarray}
Using the thermodynamic potential at $T=0$, Eq.~(\ref{OmegaT0}), we obtain
\begin{eqnarray}
\Delta=2G_{d}N_cN_f\Delta I_{\Delta},
\label{GapD}
\end{eqnarray}
and
\begin{equation}
 M-m = 2GN_c N_f M I_M\,,
 \label{GapM}
\end{equation}
where the UV divergent momentum integrals $I_{\Delta}$ and $I_M$ appearing in the two above equations are defined as
\begin{eqnarray}
I_{\Delta} & = & \int\frac{d^{3}p}{\left(2\pi\right)^{3}}
\left(\frac{1}{E_{\Delta}^{+}}+\frac{1}{E_{\Delta}^{-}}\right),
\label{Id}\\
I_{M} & = &
\int\frac{d^{3}p}{(2\pi)^{3}}\frac{1}{E}\left[\frac{E+\mu}{E_{\Delta}^{+}}
  + \frac{E-\mu}{E_{\Delta}^{-}}\right],
 \label{Im}
\end{eqnarray}
with $E = \sqrt{p^2 + M^2}$. Note that the dispersions $E_{\Delta}^s$ and $E$ are implicit or explicit functions of the quark chemical potential $\mu$. A naive introduction of a sharp cutoff $\Lambda$, as in the TRS case,
leads to the definitions of $I_M$ and $I_{\Delta}$ being expressed as
\begin{eqnarray}
I_{\Delta}^{\text{TRS}} & = & \int_0^{\Lambda}\frac{dp~p^2}{2\pi^2}
\left(\frac{1}{E_{\Delta}^{+}}+\frac{1}{E_{\Delta}^{-}}\right),
\label{IdTRS}\\
I_{M}^{\text{TRS}} & = &
\int_0^{\Lambda}\frac{dp~p^2}{2\pi^2}\frac{1}{E}\left(\frac{E+\mu}{E_{\Delta}^{+}}
  + \frac{E-\mu}{E_{\Delta}^{-}}\right),
 \label{ImTRS}
\end{eqnarray}
where medium contributions are being regularized together with the divergencies.
As already mentioned and demonstrated in many previous references~\cite{Farias:2005cr,Lopes:2025rvn,Azeredo:2024sqc,Lopes:2021tro,Avancini:2019ego,Duarte:2018kfd,Farias:2016let,Farias:2006cs}, this procedure can lead to spurious and nonphysical results, while the MSS procedure avoids these issues. In the MSS case, after separating the finite contributions from divergent integrals, $I_{\Delta}$ and $I_M$ become (see Appendix~\ref{apendix_gaps} for details)
\begin{eqnarray}
I_{\Delta}^{\text{MSS}} &=& 2I_{\text{quad}}(M_0)
-\left(\Delta^{2}-M_{0}^{2}-2\mu^2+M^{2}\right)
I{}_{\text{log}}(M_0)\nonumber\\& &+
\left[\frac{3(A^{2}+4M^{2}\mu^2)}{4}-3M_{0}^{2}\mu^{2}\right]
I_{1}+2I_{2},
 \label{IdMSS}
\end{eqnarray}
and 
\begin{eqnarray}
I_{M}^{\text{MSS}} & = & I_{\Delta}^{\text{MSS}}
-2\mu^2I_{\text{log}}(M_0)-3A\mu^{2}I_{1}+I_{3}\,,
 \label{ImMSS}
\end{eqnarray}
where $A = M_0^2 - M^2 - \mu^2 - \Delta^2$, while $I_{\text{quad}}(M_0),$  $\;I_{\text{log}}(M_0),\;I_1,\;I_2, \; I_3$
are explicitly defined in Appendix~\ref{apendix_gaps}.
In the definitions of these expressions, all medium contributions were completely removed from the UV divergent momentum integrals, with only $I_{\text{quad}}$ and $I_{\text{log}}$, which depend only on the scale $M_0$ (vacuum constituent quark mass), regularized with the three-dimensional cutoff $\Lambda$. 

%%%%%%%%%%%%%%%%%%%%%%%%%%%%%%%%%%%%%%%%%%%%%%%%%%
\subsection{Baryon density}

To evaluate the quark density in both regularization schemes, we first
observe that
\begin{eqnarray}
n & = &-\frac{\partial\Omega_{T = 0}}{\partial\mu}\nonumber\\
  & = & N_cN_f\int\frac{d^{3}p}{\left(2\pi\right)^{3}}
\left(\frac{E}{E_{\Delta}^{+}}
-\frac{E}{E_{\Delta}^{-}}\right)\nonumber\\ & &+
N_cN_f\mu\int\frac{d^{3}p}{\left(2\pi\right)^{3}}
\left(\frac{1}{E_{\Delta}^{+}}
+\frac{1}{E_{\Delta}^{-}}\right)\nonumber\\ & = & N_c
N_f\Big[I_{n}+\mu I_{\Delta}\Big],
\label{intn}
\end{eqnarray}
where in the TRS case, the UV divergent momentum integral $I_n$ in the above expression is given by
\begin{equation}
 I_n^{\text{TRS}} = \int_0^{\Lambda}\frac{dp~p^2}{2\pi^2}
 \left(\frac{E}{E_{\Delta}^{+}} -\frac{E}{E_{\Delta}^{-}}\right)\,.
 \label{InTRS}
\end{equation}
In the MSS case, we instead find the result
\begin{eqnarray}
I_{n}^{\text{MSS}} & = & -2\mu I_{\text{quad}}(M_0)\nonumber\\
& &+\mu\left(2M_{0}^{2}-5\mu^{2}-3A-2M^{2}\right)I_{\text{log}}(M_0)
\nonumber \\ & &+
\mu\left[3(M_{0}^{2}-M^2)A+5M_{0}^{2}\mu^{2}\right]I_{1}\nonumber\\
& &-
\frac{5M^{2}\mu}{4}(3A^{2}+4 M^{2}\mu^{2})I_{4}
\nonumber \\ & &+
\frac{5\mu}{4}(4M_{0}^{2}\mu^{2}-3A^{2}-8M^{2}\mu^{2})
I_{5} +I_{6},  \nonumber \\
 \label{InMSS}
\end{eqnarray}
and whose explicit derivation is shown in Appendix~\ref{apendix_gaps}, where we also give the definitions of $I_4, \, I_5$, and $I_6$.

%%%%%%%%%%%%%%%%%%%%%%%%%%%%%%%%%%%%%%%%%%%%%%%%%%
\section{Numerical Results}
\label{sec3}

In this section, we show the numerical results for both physical pion mass and heavy pion mass used in the LQCD simulations that we have considered in our comparison.
This section is divided into two subsections.
In the first one, we show the numerical results for the NJL model with a parametrization using the physical pion mass, similar to other approaches in effective theories of two-color QCD \cite{Ratti:2004ra}.
Then, we show the results for the heavy pion parametrization and compare the obtained speed of sound with the lattice data from Ref.~\cite{Iida:2024irv}.

%%%%%%%%%%%%%%%%%%%%%%%%%%%%%%%%%%%
\subsection{Physical pion results}

Let us first start by showing how the TRS and MSS approaches produce results that can be fundamentally different.

%%%%%%%%%%FIG 1%%%%%%%%%%%%%%%%%%
\begin{figure}[!htb]
\includegraphics[width=0.45\textwidth]{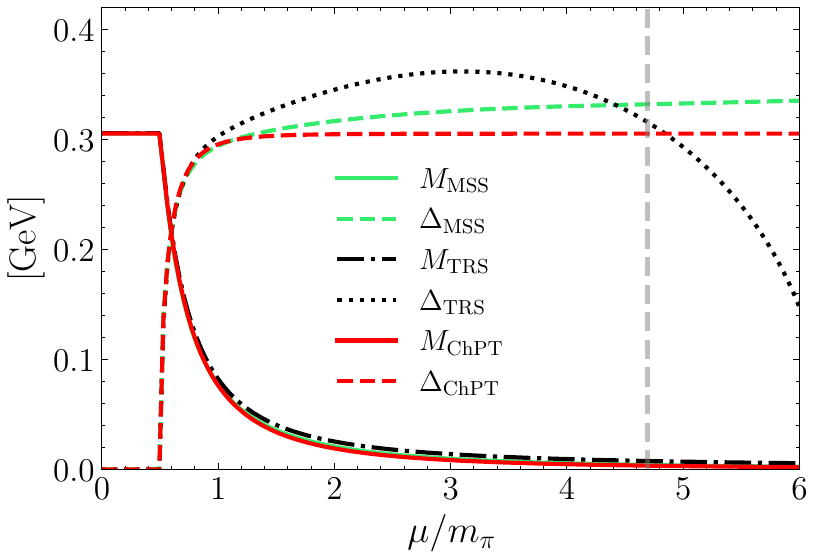}
\caption{\label{fig1} Effective quark mass $M$ (in units of GeV) results for ChPT (solid red line), TRS (black dash-dotted line), and for MSS (solid green line).
The results for the diquark condensate $\Delta$ (in units of GeV) for ChPT (dashed red line), TRS (dotted black line), and for MSS (dashed green line). The gray dashed vertical line marks the model energy scale, $\Lambda$. All quantities are expressed as
      functions of the quark chemical potential divided by the pion mass,
      $\mu/m_\pi$.}
\end{figure}

In {}Fig.~\ref{fig1}, we show the results for the effective quark mass $M$ and the diquark condensate $\Delta$ as a function of the baryon chemical potential scaled by the pion mass. The TRS and MSS results are also compared with the ChPT predictions for the effective quark mass $M_\text{ChPT}$ and the diquark condensate $\Delta_\text{ChPT}$ summarized in Table~\ref{tabMD}. All ChPT values were computed using the parameter sets corresponding to each case discussed in this work, as presented in Table~\ref{tab:params}

\begin{table}[b]
    \caption{Effective quark mass $M$ and diquark condensate $\Delta$ as functions of the baryon chemical potential adapted from~\cite{Kogut:2000ek,Duarte:2015ppa}.\label{tabMD}}
    \begin{ruledtabular}
    \begin{tabular}{ccc}
    &
    \multicolumn{1}{c}{$\mu < m_\pi/2$}&
    \multicolumn{1}{c}{$\mu > m_\pi/2$}\\
        \hline
        $M_\text{ChPT}$&$\displaystyle{M_0}$&$\displaystyle{M_0\left(\frac{m_\pi}{2\mu}\right)^2}$\\
        $\Delta_\text{ChPT}$&$\displaystyle{0}$&$\displaystyle{M_0\sqrt{1-\left(\frac{m_\pi}{2\mu}\right)^4}}$\\
    \end{tabular}
    \end{ruledtabular}
\end{table}

{}From the results shown in  {}Fig.~\ref{fig1}, it can be observed that, for higher values of $\mu$, the value of $\Delta$ begins to decrease when using the TRS approach, eventually reaching zero at $\mu/m_{\pi}\sim 6.2 \ \ (\mu \approx 0.86$ GeV). An important aspect to highlight is that, since the energy scale for the model is $\Lambda$, shown as a gray vertical dashed line, it can be argued that the results may no longer be valid for such high values of $\mu$. Considering that $\Lambda = 657$ MeV, the results are reliable up to $\mu \sim 5m_{\pi}$. In this range, the TRS curve has already begun to decrease, while the MSS curve continues to increase slightly, which is a result consistent with the ChPT predictions.

\begin{figure}[!htb]
    \includegraphics[width=0.47\textwidth]{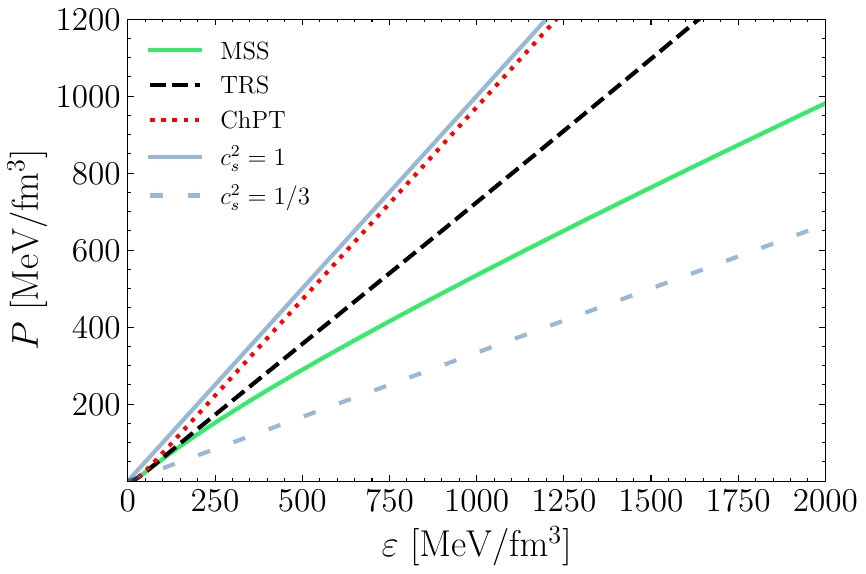}
    \caption{\label{fig2} The pressure density $P$ as a function of energy density $\varepsilon$ (all in units of MeV/fm$^3$), for the MSS (continuous green line), TRS
      (dashed black line) and ChPT (dotted red line).  The solid and dashed gray lines correspond to the values of constant speed of sound EoS, $c_s^2=1$ and $c_s^2=1/3$, respectively.}
\end{figure}

We next move to the primary objective of this work, which is to show that the peak structure in the sound velocity may be reproduced by separating the finite contributions from divergent integrals and regularizing only the vacuum in the MSS framework, while this peak structure is absent in the TRS case. To this end, we first show in {}Fig.~\ref{fig2} the behavior of the EoS.
The ChPT expressions are shown in Table~\ref{tab2} and were extracted from Ref.~\cite{Hands:2006ve}. All parameter values used in our analysis were again obtained from Table~\ref{tab:params}.

\begin{table}[b]
    \caption{Thermodynamic quantities in the ChPT approach from Ref.~\cite{Hands:2006ve}.\label{tab2}}
    \begin{ruledtabular}
    \begin{tabular}{ccc}
    &
    \multicolumn{1}{c}{$2\mu < m_\pi$}&
    \multicolumn{1}{c}{$2\mu > m_\pi$}\\
        \hline
        $n_{B,\text{ChPT}}$ & $0$ & $\displaystyle{N_f f_{\pi}^2\mu\left(1 - \frac{m_{\pi}^4}{\mu^4} \right)}$ \\
        $P_{\text{ChPT}}$ & $0$ & $\displaystyle{N_f f_{\pi}^2\left(\mu^2 + \frac{m_{\pi}^4}{\mu^2} - 2m_{\pi}^2\right)}$ \\
        $\varepsilon_{\text{ChPT}}$ & $0$ & $\displaystyle{N_f f_{\pi}^2\left(\mu^2 - 3\frac{m_{\pi}^4}{\mu^2} + 2m_{\pi}^2\right)}$ \\
        ${c_s^2,}_{\text{ChPT}}$ & 0 & $\displaystyle{\left(1-\frac{m_\pi^4}{\mu^4}\right)\Bigg/\left(1+\frac{3 m_\pi^4}{\mu^4}\right)}$ \\
    \end{tabular}
    \end{ruledtabular}
\end{table}

%%%%%%%%%%%%%%Fig 2%%%%%%%%%%%%%%%%%%%%%%

The numerical results shown in {}Fig.~\ref{fig2} indicates that in the regime of low baryon chemical potential (or equivalently, for low densities), the curves for both TRS and MSS cases are very similar, but begin to differ as the density increases. This behavior is expected for the following reasons. As shown in {}Fig~\ref{fig1}, $\Delta = 0$ for $\mu\leq m_{\pi}/2$, and the masses exhibit very similar behavior for both schemes. Up to $\mu \sim m_{\pi}$, the results for $M$ and $\Delta$ in the TRS and MSS cases are also close to each other, and beyond this point, the differences are expected to be more pronounced. This is reflected in all thermodynamic quantities but is most evident in the energy density. The difference can also be seen in the pressure separation values $P$, as a function of the energy density $\varepsilon$, and shown in {}Fig.~\ref{fig2}. Both curves begin to differ for $\varepsilon \approx 100$ MeV/fm$^3$, where the MSS EoS becomes softer as the density increases. {}From this value onward, we can notice a decrease in the value of the speed of sound, corroborating the hypothesis that the thermodynamic quantities are sensitive to the regularization procedure.

%%%%%%%%%%%%%%%%Fig 3%%%%%%%%%%%%%%%%%%%%%%%%
\begin{figure}[!htb]
    % \flushleft
\includegraphics[width=0.47\textwidth]{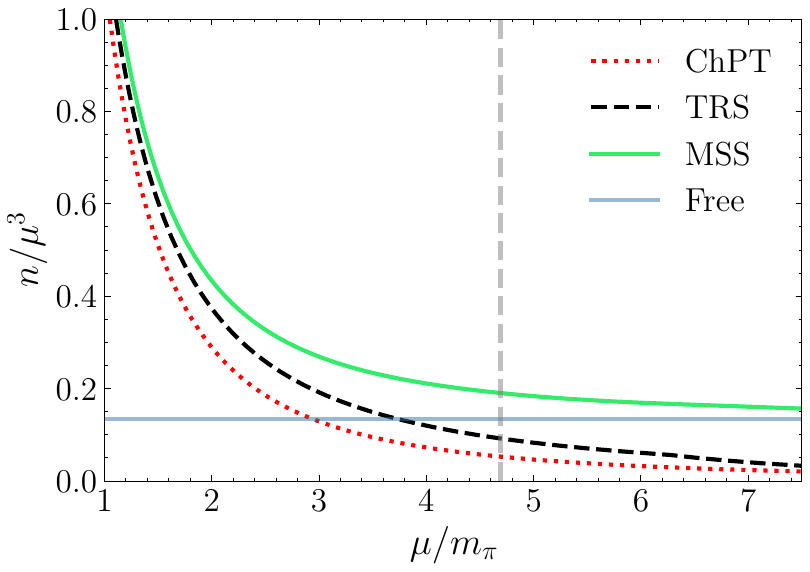}
    \caption{\label{fig3} The baryon number density divided by baryonic chemical potential to the third power, as a function of
      baryon chemical potential normalized by the pion mass,
      $\mu/m_\pi$, for ChPT (red dotted line), TRS (black dashed line), MSS (green solid line), and the massless free fermion gas (blue solid line). The gray dashed vertical line marks the model energy scale, $\Lambda$.}
\end{figure}

{}Figure~\ref{fig3} further supports the above interpretation of the results. It shows, for instance, the results for particle number density divided by $\mu^3$.
The results 
indicate that both TRS and ChPT curves for $n/\mu^3$ fail to converge to the conformal limit as $\mu/m_\pi$ increases way past the onset of diquark condensation into the superfluid phase. 
Moreover, ChPT and TRS curves for $n/\mu^3$ cross the free gas value near $\mu = 3  m_\pi$ and $\mu = 4 m_\pi$, respectively, continuing to decrease, reaching even lower values at $\Lambda$, indicated by the gray vertical dashed line in Fig.~\ref{fig3}.
In opposition to these results, the MSS converges to the conformal limit for high $\mu/m_\pi$, at first, concomitant to the TRS curve until near $\mu = 1.2 m_\pi$.

%%%%%%%%%%%%%%Fig 4%%%%%%%%%%%%%%%%%%%%%%%%%%%%
\begin{figure}[!htb]
\includegraphics[width=0.47\textwidth]{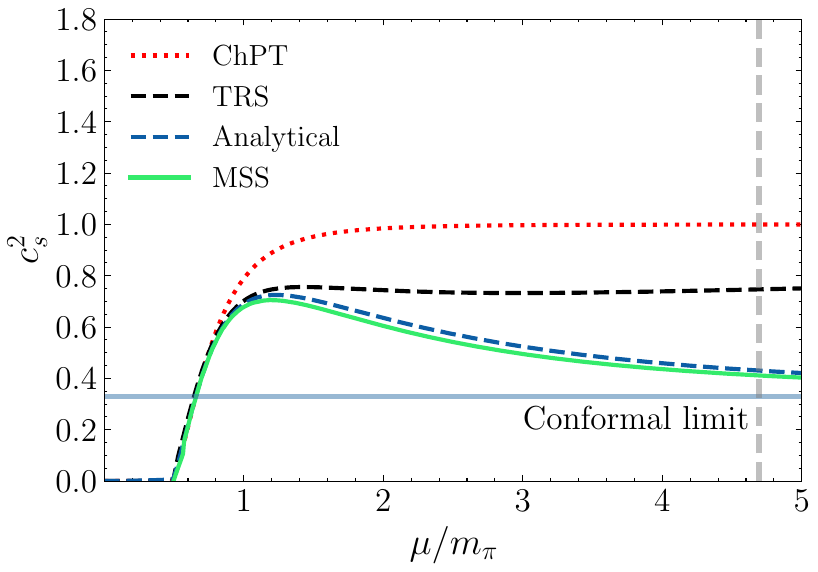}
    \caption{\label{fig4} Speed of sound squared, $c_s^2$, as a function of
      quark chemical potential normalized by the pion mass,
      $\mu/m_\pi$, for ChPT (red dotted line), TRS (black dashed line), the analytical expression Eq.~(\ref{fit}) (blue dot-dashed line), and MSS (green solid line). The gray dashed vertical line marks the model energy scale, $\Lambda$.}
\end{figure}

Both ChPT (red dotted curve) and TRS (black dashed curve) in {}Fig.~\ref{fig3} yield values of $n/\mu^3$ that are lower than those of a free fermion gas for $\mu > 4m_\pi$.
Meanwhile, for MSS (solid green curve), the behavior of $n/\mu^3$ converges monotonically to that of a free fermion gas as $\mu$ increases. An important point is that our results can be trusted within the validity range of the model, $\mu \approx \Lambda$. However, when extrapolating to higher values of $\mu$, we observe that among the procedures considered in this study, MSS is the only one that reproduces the expected Stefan-Boltzmann limit for $n/\mu^3$.

The speed of sound squared is shown in {}Fig.~\ref{fig4}.
This is one of the most important observables as far as lattice QCD results are concerned.
The MSS curve (solid green line) reproduces the peak behavior observed in lattice QCD simulations, while TRS fails to do so. 
This confirms that the softening of the MSS EoS at high densities is essential to capture the correct structure of $c_s^2$. 

Here, we note that the MSS result for $c_s^2$ ~is well approximated by the analytical function given by
\begin{equation}
c_s^2 \simeq \frac{2 \Delta^2 + \mu^2}{2 N_c N_f M^2+ 2 \Delta^2+ 3\mu^2}.
\label{fit}
\end{equation}
The curve for this analytical function of $c_s^2$ is displayed alongside the ChPT, TRS, and MSS curves in {}Fig.~\ref{fig4}. 
Equation~(\ref{fit}) is similar to the expression found in Ref.~\cite{Gholami:2025afm} for large densities and derived in that reference
under renormalization group techniques.
Since the peak in $c_s^2$ is also required to satisfy astrophysical constraints from gravitational-wave data, the MSS emerges as a physically reliable regularization procedure.

Next, we will explicitly make use of a parametrization that matches the heavy pion mass utilized in two color QCD lattice setups. 

%%%%%%%%%%%%%%%%%%%%%%%%%%%%%%%%%%%%%%%%%%%%%%%%%%
\subsection{Heavy pion results}

\begin{figure}[!htb]
\includegraphics[width=0.45\textwidth]{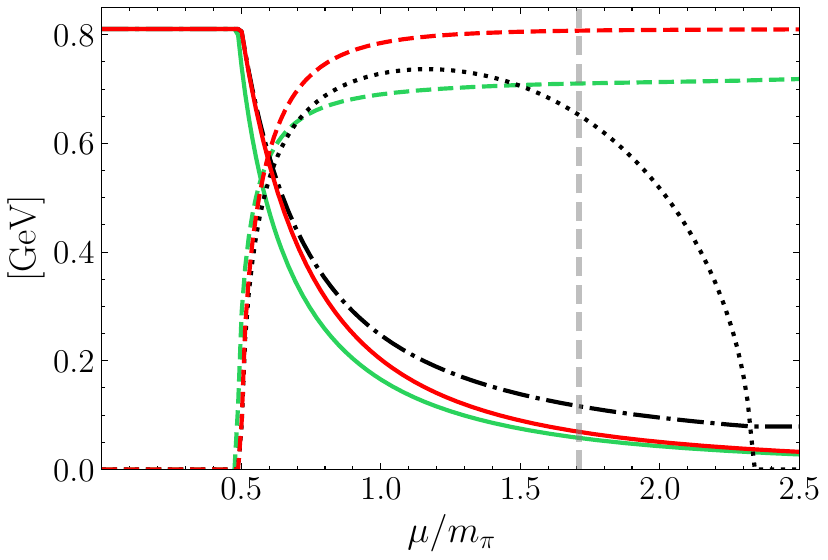}% Here is how to import EPS art
\caption{\label{fig5} Effective quark mass $M$ (in units of GeV) results for ChPT (solid red line), TRS (black dash-dotted line), and for MSS (solid green line).
      The results for the diquark condensate $\Delta$ (in units of GeV) for ChPT (dashed red line), TRS (dotted black line) and for MSS (dashed green line). The gray dashed vertical line marks the model energy scale, $\Lambda$. All quantities are expressed as
      functions of the quark chemical potential divided by the pion mass,
      $\mu/m_\pi$.}
\end{figure}

We begin by showing the order parameters for the heavy pion parametrization in Fig. \ref{fig5}, as well as the upper limit $\Lambda$ represented as a gray vertical dashed line around $\mu \approx 1.7 m_\pi$.
The heavy pion parameters are given in Table~\ref{tab:params}.

%%%%%%%%%%%%%%Fig 5%%%%%%%%%%%%%%%%%%%%%%%%%%%%

In Fig. \ref{fig5} we can see that the value for the diquark condensate in the TRS scheme, $\Delta_{\text{TRS}}$, finds a maximum at $\mu \approx 1.2 m_\pi$, decreasing afterward. 
The value of $\Delta_{\text{TRS}}$ keeps decreasing even after $\Delta_{\text{TRS}}$ crosses the $\mu = \Lambda$ limit and reaches zero right above $\mu = 2.3 m_\pi$ ($\mu = 1.697$ GeV). 
Meanwhile, both values for $\Delta_{\text{MSS}}$ and $\Delta_{\text{ChPT}}$ monotonically increase, reaching a plateau for values of $\mu > \Lambda$, where the plateau for $\Delta_{\text{MSS}}$ sits $0.1$ GeV lower than the one for $\Delta_{\text{ChPT}}$. 
The effective masses follow the same qualitative behavior for all the methods, with the difference that $M_{\text{TRS}}$ converges to a value slightly higher than both $M_{\text{MSS}}$ and $M_{\text{ChPT}}$.  
Although all curves for diquark condensate are arguably close to each other until $\mu = \Lambda$, with $\Delta_{\text{TRS}}$ staying close to $\Delta_{\text{MSS}}$ until this limit, the qualitative behavior of $\Delta_{\text{TRS}}$ heavily influences the thermodynamics of quark matter, as we will see next.

\begin{figure}[!htb]
\includegraphics[width=0.47\textwidth]{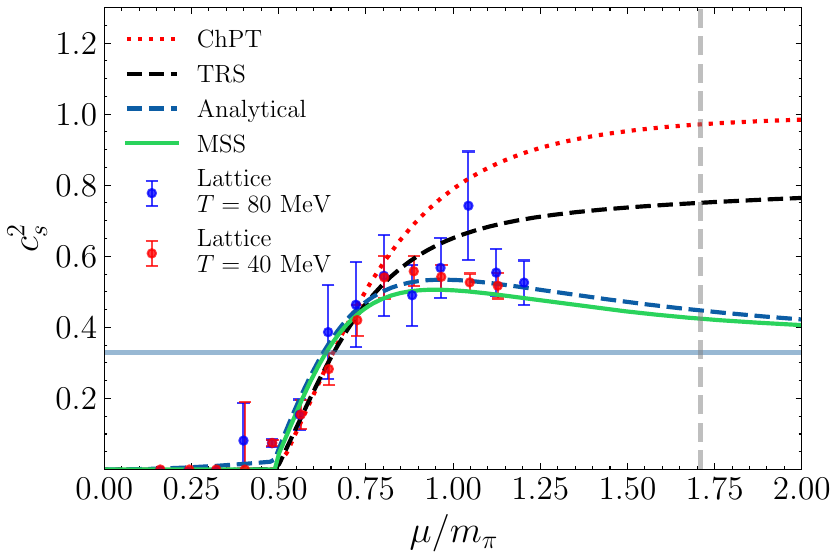}
    \caption{\label{fig6} Speed of sound squared, $c_s^2$, as a function of
      quark chemical potential normalized by the pion mass,
      $\mu/m_\pi$, for ChPT (red dotted line), TRS (black dashed line), the analytical expression Eq.~(\ref{fit}) (blue dot-dashed line), and MSS (green solid line). The gray dashed vertical line marks the model energy scale, $\Lambda$. Blue (red) dots with error bars indicate the lattice datapoints for $T=80$ MeV ($T=40$ MeV) adapted from Ref.~\cite{Iida:2024irv}.}
\end{figure}

We now follow for the results on the speed of sound for heavy pion parametrization of the NJL model in both TRS and MSS compared with lattice results from \cite{Iida:2024irv} and ChPT, shown in Fig. \ref{fig6}.
Here we chose to compare our data with two sets of datapoints for speed of sound squared from LQCD with two different temperatures, aiming to approach the zero temperature limit of two color QCD.
One set was obtained setting the temperature as $T=80$ MeV and shown in blue with the corresponding error bars. 
The other one was obtained by setting the temperature to $T=40$ MeV and is shown in red. 

%%%%%%%%%%%%%%Fig 6%%%%%%%%%%%%%%%%%%%%%%%%%%%%

In Fig. \ref{fig6}, one may see the same qualitative behavior for the speed of sound squared seen in Fig. \ref{fig4} across all methods.
Here, the $c_s^2$ curve obtained via MSS best fits the lattice data, showing the clear peak structure around $\mu = m_\pi$, where its curvature in the interval $m_\pi< \mu < \Lambda$ indicates a steady convergence toward the conformal limit at $c_s^2 =1/3$ from above.
Meanwhile, the TRS and ChPT curves for speed of sound squared overshoot the peak and grow monotonically to $c_s^2 = 1$ in ChPT and $c_s^2 \approx 0.8$ in TRS.
Another feature of the MSS procedure is the consistent extrapolation of thermodynamic quantities for $\mu >\Lambda$, where other methods fail to reproduce free fermion gas limits at high density.

%%%%%%%%%%%%%%%%%%%%%%%%%%%%%%%%%%
\section{Final Remarks}
\label{conclusions}

In this paper, we have presented a comprehensive analysis of the two-color SU(2) NJL model within the MSS at zero temperature and finite density. We have focused on the chiral symmetry breaking, studying the dynamically generated quark mass, the diquark condensate, and the analysis of the thermodynamic properties for two different parametrizations.
The choice of $N_c = 2$ is particularly advantageous because QC$_2$D does not suffer from the sign problem, allowing direct comparisons with lattice QCD. This, in turn, provides a unique opportunity to test the validity of effective models such as the NJL, ensuring that predictions can be compared against first-principles calculations.
In order to compare the NJL results with lattice QCD, we considered another parametrization set alongside with the usual one with a physical pion mass. 

In addition to chiral symmetry, this study explicitly considers the behavior of the diquark condensate. In QC$_2$D, Cooper pairs form color singlets, so the system represents a true superfluid in which only the global baryon number symmetry is spontaneously broken; this feature distinguishes the QC$_2$D theory from QC$_3$D and highlights the relevance of investigating both chiral and diquark condensates in dense matter.

Our results demonstrate that the MSS provides a consistent and physically reliable framework by properly separating vacuum and medium contributions, avoiding spurious and nonphysical medium effects that can arise in conventional regularization schemes. The dynamically generated quark mass and diquark condensate both exhibit smooth and stable behavior as functions of the chemical potential, indicating that MSS captures the essential features of chiral symmetry breaking and superfluid formation in dense quark matter. 

Thermodynamic observables derived from MSS, such as pressure, energy density, and especially the squared speed of sound $c_s^2$, show physically consistent trends; in particular, $c_s^2$ behaves in a way that aligns more closely with QCD expectations, avoiding nonphysical results often observed in other schemes. Furthermore, other thermodynamic quantities computed with MSS tend to approach lattice QCD results at low temperature, confirming that the two-color NJL model with MSS can quantitatively reproduce key features of strongly interacting matter.
Our results, alongside previous ones, corroborate the hypothesis that careless UV regularization of medium-dependent quantities has significant effects on the thermodynamics of nonrenormalizable theories.
The MSS procedure separates the finite medium contributions from the vacuum contributions, where the UV regularization must be applied.
Thus, the MSS procedure is not just another UV regularization procedure, but a procedure required to properly account for medium-dependent terms in UV quantities.

Looking ahead, an important perspective is the extension of this framework to finite temperatures, providing a direct way to compare MSS predictions with state-of-the-art lattice QCD data. These comparisons would not only offer a robust benchmark for the model and the chosen regularization scheme but also improve our understanding of the QCD phase diagram, especially in regions where temperature and density are relevant.

In conclusion, the study conducted in this paper establishes the MSS as a robust and reliable approach to investigate dense quark matter in the SU(2) NJL model at zero temperature, capturing both chiral symmetry breaking and superfluidity through diquark condensation. This work, therefore, lays a solid foundation for future studies of the QCD equation of state, superfluidity, and phase structure in dense and hot environments within the two-color NJL framework, while emphasizing that $N_c = 2$ allows direct quantitative tests against first-principles lattice QCD results.

%%%%%%%%%%%%%%%%%%%%%%%%%%%%%%%%%%%%%%%%%%%%%%%%
\begin{acknowledgments}
This work was partially supported by Conselho Nacional de
Desenvolvimento Cient\'{\i}fico e Tecnol\'ogico (CNPq), Grants No. 312032/2023-4, 402963/2024-5 and 445182/2024-5 (R.L.S.F.), 307286/2021-5 (R.O.R.); Funda\c{c}\~ao
de Amparo \`a Pesquisa do Estado do Rio Grande do Sul (FAPERGS),
Grants Nos. 24/2551-0001285-0 (R.L.S.F.) and
23/2551-0000791-6 and 23/2551-0001591-9 (D.C.D.); Funda\c{c}\~ao
Carlos Chagas Filho de Amparo \`a Pesquisa do Estado do Rio de Janeiro
(FAPERJ), Grants No. E-26/201.150/2021 (R.O.R.); Coordena\c{c}\~ao de
Aperfei\c{c}oamento de Pessoal de N\'ivel Superior - Brasil (CAPES) -
Finance Code 001 (A.E.B.P). The work is also part of the project
Instituto Nacional de Ci\^encia e Tecnologia - F\'isica Nuclear e
Aplica\c{c}\~oes (INCT - FNA), Grants No. 464898/2014-5 and No. 408419/2024-5, and supported
by the Ser\-ra\-pi\-lhei\-ra Institute (Grant No. Serra -
2211-42230). R. L. S. F. acknowledges
the kind hospitality of the Center for Nuclear Research at Kent State University, where part of this work was
done.   
\end{acknowledgments}

\section{Data Avaliability}

The
data that support the findings of this article are not publicly available. The data are available from the authors upon reasonable request.

%%%%%%%%%%%%%%%%%%%%%%%%%%%%%%%%%%%%%%%%%%
\appendix
\section{THE GAP EQUATIONS FOR $\Delta$ AND $M$ IN THE MSS} 
\label{apendix_gaps}

In Sec.~\ref{gaps}, we presented the expressions for the gap equations for $\Delta$ and $M$. These are obtained by minimizing the thermodynamic potential with respect to these quantities. 

%%%%%%%%%%%%%%%%%%%%%%%%%%%%%%%%%%%%%%%%%%
%% 
\subsection{Gap equation for $\Delta$ in the MSS}

The gap equation for $\Delta$ in the TRS implementation consists of introducing a three-dimensional cutoff $\Lambda$ to regularize all UV divergent momentum integrals. Conversely, for MSS we begin by rewriting~\eqref{Id} as 
\begin{eqnarray}
I_{\Delta} & = &
\sum_{s=\pm1}\int\frac{d^{3}p}{\left(2\pi\right)^{3}}\frac{1}{E_{\Delta}^{s}}\nonumber\\ &
= &\frac{1}{\pi}\sum_{s=\pm1}
\int_{-\infty}^{+\infty}dp_{4}\int\frac{d^{3}p}{\left(2\pi\right)^{3}}
\frac{1}{p_{4}^{2}+(E_{\Delta}^{s})^{2}}\,,
 \label{tempIvD}
\end{eqnarray}
such that 
\begin{eqnarray}
& &\frac{1}{2}
  \sum_{s=\pm1}\int\frac{d^{3}p}{\left(2\pi\right)^{3}}
  \frac{1}{E_{\Delta}^{s}} \nonumber \\& &=\sum_{s=\pm1}\int_{-\infty}^{+\infty}
\frac{dp_{4}}{2\pi}\int\frac{d^{3}p}{\left(2\pi\right)^{3}}
\frac{1}{p_{4}^{2}+(E_{\Delta}^{s})^{2}}.
\label{identD4k}
\end{eqnarray}
By iterating two times using the identity
\begin{eqnarray}
& &\frac{1}{p_{4}^{2}+\left(E+s\mu\right)^{2}+\Delta^{2}} \nonumber\nonumber\\ & &= \frac{1}{p_{4}^{2}+p^{2}+M_{0}^{2}} \nonumber\\& &\ \ \ \ -
\frac{\mu^{2}+2sE\mu+\Delta^{2}+M^{2}-M_{0}^{2}}{\left(p_{4}^{2}+p^{2}+
  M_{0}^{2}\right) \left[p_{4}^{2}
+\left(E+s\mu\right)^{2}+\Delta^{2}\right]},
 \label{ident}
\end{eqnarray}
we obtain
\begin{eqnarray}
 & &\frac{1}{p_{4}^{2}
+\left(E+s\mu\right)^{2}+\Delta^{2}}\nonumber\\
& &= \frac{1}{p_{4}^{2}+p^{2}+M_{0}^{2}}-\frac{2sE\mu-A}{\left(p_{4}^{2}+p^{2}+M_{0}^{2}\right)^{2}}+
\frac{\left(2sE\mu-A\right)^{2}}{\left(p_{4}^{2}+p^{2}+M_{0}^{2}\right)^{3}}\nonumber \\& & \ \ \ \ \ -
\frac{\left(2sE\mu-A\right)^{3}}{\left(p_{4}^{2}+
  p^{2}+M_{0}^{2}\right)^{3}\left[p_{4}^{2}
+\left(E+s\mu\right)^{2}+\Delta^{2}\right]},
 \label{r3it}
\end{eqnarray}
where $A=M_{0}^{2}-M^{2}-\mu^{2}-\Delta^{2}$. After some straightforward 
 algebraic manipulations, performing the summation over $s$ and the integration over the momentum in the $p_{4}$ (Euclidean direction), as outlined in Eq.~(\ref{identD4k}),
we obtain the result
\begin{eqnarray}
 & &\sum_{s=\pm1}\int_{-\infty}^{+\infty}\frac{dp_{4}}{2\pi}
  \int\frac{d^{3}p}{\left(2\pi\right)^{3}}\frac{1}{p_{4}^{2}+(E_{\Delta}^{\pm})^{2}}
  \nonumber\\& &=
I_{\text{quad}}(M_0)-\frac{\left(\Delta^{2}-M_{0}^{2}-2\mu^{2}+M^{2}\right)}{2}I_{\text{log}}(M_0)
\nonumber\\ & & \ \ \ \ \ +
\left[\frac{3\left(A^{2}+4M^{2}\mu^{2}\right)}{8}
  -\frac{3\mu^{2}M_{0}^{2}}{2}\right]I_{\text{1}} +  I_{\text{2}},
 \label{temp4}
\end{eqnarray}
where we have defined the quantities:
\begin{eqnarray}
I_{\text{quad}}(M_0) & = &
\int\frac{d^{3}p}{\left(2\pi\right)^{3}}\frac{1}{\sqrt{p^{2}+M_{0}^{2}}},
\label{quad}
\\ I_{\text{log}}(M_0) & = & \int\frac{d^{3}p}{\left(2\pi\right)^{3}}
\frac{1}{\left(p^{2}+M_{0}^{2}\right)^{\frac{3}{2}}},
\label{log}
\\ I_{\text{1}} & = & \int\frac{d^{3}p}{\left(2\pi\right)^{3}}
\frac{1}{\left(p^{2}+M_{0}^{2}\right)^{\frac{5}{2}}},
\label{Ifin1}
\\ I_{\text{2}} & = &
\frac{15}{32}\sum_{s=\pm1}\int\frac{d^{3}p}{\left(2\pi\right)^{3}}
\int_{0}^{1} dx (1-x)^{2}\nonumber\\& & \times
\frac{(A-2sE\mu)^{3}}
     {\left[(2sE\mu-A)x+p^{2}+M_{0}^{2}\right]^{\frac{7}{2}}}.
 \label{Ifin2}
\end{eqnarray}

Comparing Eq.~(\ref{tempIvD}) and the left-hand side of
Eq.~(\ref{identD4k}), we can see that
\begin{equation}
\frac{1}{2}I_{\Delta}=\sum_{s=\pm1}\int_{-\infty}^{+\infty}\frac{dp_{4}}{2\pi}
\int\frac{d^{3}p}{\left(2\pi\right)^{3}}\frac{1}{p_{4}^{2}+
  (E_{\Delta}^{\pm})^{2}},
\end{equation}
and, therefore,
\begin{eqnarray}
I_{\Delta}^{\text{MSS}} &=& 2I_{\text{quad}}(M_0)
-\left(\Delta^{2}-M_{0}^{2}-2\mu^{2}+M^{2}\right)
I{}_{\text{log}}(M_0)+2I_{\text{fin,2}}\nonumber \\ & & +
\left[\frac{3(A^{2}+4M^{2}\mu^{2})}{4}-3M_{0}^{2}\mu^{2}\right]
I_{\text{fin,1}}.
 \label{IvrDfinal}
\end{eqnarray}

%%%%%%%%%%%%%%%%%%%%%%%%%%%%%%%%%%%%%%%%%%
%% 
\subsection{Gap equation for $M$ in the MSS}

{}For the mass gap equation in MSS, we first write 
\begin{eqnarray}
I_{M} & = & \sum_{s=\pm1}\int\frac{d^{3}p}{(2\pi)^{3}}\frac{1}{E}
\frac{E+s\mu}{E_{\Delta}^{s}} \nonumber \\ &=&
\sum_{s=\pm1}\int\frac{d^{3}p}{(2\pi)^{3}} \frac{1}{E_{\Delta}^{s}}\nonumber\\& & +
\sum_{s=\pm1}\int\frac{d^{3}p}{(2\pi)^{3}}\frac{s\mu}{E}
\int_{-\infty}^{\infty}\frac{dp_{4}}{\pi}\frac{1}{p_{4}^{2}+(E_{\Delta}^{s})^{2}}.
\end{eqnarray}
Note that the first momentum integral on the right-hand side of the above equation is exactly $I_{\Delta}$. To deal
with the second term, we can use
the result Eq.~(\ref{r3it}) and write, after performing the integration in $p_{4}$,
\begin{eqnarray}
  & &\frac{\mu}{E}\int_{-\infty}^{\infty}\frac{dp_{4}}{\pi}\sum_{s=\pm1}
  \frac{s}{p_{4}^{2}+\left(E+s\mu\right)^{2}+\Delta^{2}} \nonumber\\ & &=\sum_{s=\pm1}\frac{s\mu\left(A-2sE\mu\right)^{3}}
    {E\left(p_{4}^{2}+p^{2}+M_{0}^{2}\right)^{3}\left[p_{4}^{2}
        +\left(E+s\mu\right)^{2}+\Delta^{2}\right]}\nonumber\\& & \ \ \ \
-\frac{4\mu^{2}}{\left(p_{4}^{2}+p^{2}+M_{0}^{2}\right)^{2}}
-\frac{8A\mu^{2}}{\left(p_{4}^{2}+p^{2}+M_{0}^{2}\right)^{3}}.  
 \label{tempI7}
\end{eqnarray}
Then, we obtain
\begin{eqnarray}
  & &\int\frac{d^{3}p}{(2\pi)^{3}}\frac{\mu}{E}\int_{-\infty}^{\infty}
  \frac{dp_{4}}{\pi}
  \sum_{s=\pm1}\frac{s}{p_{4}^{2}+\left(E+s\mu\right)^{2}+\Delta^{2}} \nonumber\\
& &= -2\mu^{2}I_{\text{log}}(M_0) - 3A\mu^{2}I_{\text{1}} + I_{\text{3}},
\label{intImb}
\end{eqnarray}
with
\begin{eqnarray}
I_{\text{3}}
  &=&\frac{15}{16}\sum_{s=\pm1}\int\frac{d^{3}p}{(2\pi)^{3}}\int_{0}^{\infty}
  dt  \frac{t^2}{\sqrt{1+t}}\nonumber\\& &\times
\frac{s\mu\left(A-2sE\mu\right)^{3}}{E
\left[\left(p^{2}+M_{0}^{2}\right)t+\left(E+s\mu\right)^{2}+
    \Delta^{2}\right]^{\frac{7}{2}}},
\end{eqnarray}
where we have used the {}Feynman parametrization formula,
\begin{equation}
\frac{1}{a^{m}b^{n}}=\frac{\Gamma(m+n)}{\Gamma(m)\Gamma(n)}\int_{0}^{\infty}
\frac{t^{m-1}dt}{(at+b)^{m+n}}.
\label{FeynParam}
\end{equation}
{}Finally, the $I_M$ integral in the MSS framework becomes
\begin{eqnarray}
I_{M}^{\text{MSS}} & = & I_{\Delta}^{\text{MSS}}
-2\mu^{2}I_{\text{log}}(M_0)-3A\mu^{2}I_{\text{1}}+I_{\text{3}}\,.
 \label{ResIm}
\end{eqnarray}

While $I_{\Delta}$ and $I_M$ exhibit quadratic and logarithmic UV divergencies, the momentum integral associated with the quark density (or baryon density), $I_{n}$, presents a cubic UV divergency. In this context, to separate the divergent integrals from the medium contributions, one needs to apply the identity~\eqref{ident} three times. We start from:
\begin{eqnarray}
I_n &=&
\sum_{s=\pm1}\int\frac{d^{3}p}{(2\pi)^{3}}\frac{sE}{E_{\Delta}^{s}}
\nonumber \\ & =&
\int\frac{d^{3}p}{(2\pi)^{3}}E\sum_{s=\pm1}s\left[\frac{1}{\pi}
  \int_{-\infty}^{+\infty}\frac{dp_{4}}{p_{4}^{2}+
\left(E_{\Delta}^{s}\right)^{2}}\right].
\label{T1d4}
\end{eqnarray}
After the three iterations of ~\eqref{ident}, we get 
\begin{eqnarray}
& &\frac{1}{p_{4}^{2}+\left(E+s\mu\right)^{2}+\Delta^{2}}\nonumber\\
& & = \frac{1}{p_{4}^{2}+p^{2}+M_{0}^{2}}
+\frac{A-2sE\mu}{\left(p_{4}^{2}+p^{2}+M_{0}^{2}\right)^{2}}
\nonumber\\& & \ \ \ \ \ +
\frac{\left(A-2sE\mu\right)^{2}}{\left(p_{4}^{2}+p^{2}+M_{0}^{2}\right)^{3}}
 +\frac{\left(A-2sE\mu\right)^{3}}
{\left(p_{4}^{2}+p^{2}+M_{0}^{2}\right)^{4}}\nonumber\\& & \ \ \ \ \  +
\frac{\left(A-2sE\mu\right)^{4}}{\left(p_{4}^{2}+p^{2}+M_{0}^{2}\right)^{4}
\left[p_{4}^{2}+\left(E+s\mu\right)^{2}+\Delta^{2}\right]}.
\label{r4it}
\end{eqnarray}
After some straightforward
algebraic manipulations, performing the summation over $s$ and the
integrations over $p_{4}$ as indicated in Eq.~(\ref{T1d4}), we obtain 
\begin{eqnarray}
I_{n}^{\text{MSS}} & = & -2\mu I_{\text{quad}}(M_0)
+\mu\Big(2M_{0}^{2}-5\mu^{2}-3A-2M^{2}\Big)I_{\text{log}}(M_0)\nonumber\\& & +
\mu\Big[3(M_{0}^{2}-M^2)A+5M_{0}^{2}\mu^{2}\Big]I_{\text{1}}
\nonumber \\ & &-
\frac{5M^{2}\mu}{4}(3A^{2}+4M^{2}\mu^{2})I_{\text{4}}
\nonumber\\& & +\frac{5\mu}{4}(4M_{0}^{2}\mu^{2}-3A^{2}-8M^{2}\mu^{2})
I_{\text{5}} +I_{\text{6}},
 \label{rI8}
\end{eqnarray}
where $I_{\rm quad}, \, I_{\rm log}$ and $I_{1}$ were previously defined, while  $I_{4}, \, I_{5}$, and $I_{6}$ are given by
\begin{eqnarray}
I_{\text{4}} & = & \int\frac{d^{3}p}{(2\pi)^{3}}
\frac{1}{\left(p^{2}+M_{0}^{2}\right)^{\frac{7}{2}}}, 
\\ I_{\text{5}} & = & \int\frac{d^{3}p}{(2\pi)^{3}}\frac{p^2}{\left(p^{2}+M_{0}^{2}\right)^{\frac{7}{2}}}, 
\\ I_{\text{6}} & = &
\frac{35}{32}\sum_{s=\pm1}\int\frac{d^{3}p}{(2\pi)^{3}}
\int_{0}^{\infty}dt \nonumber\\\times& &\frac{t^3}{\sqrt{1+t}} 
\frac{sE(A-2sE\mu)^{4}}{
  \left[(p^{2}+M_{0}^{2})t+(E+s\mu)^{2}+\Delta^{2}\right]^{\frac{9}{2}}}.
\end{eqnarray}
\vspace{1cm}
%%%%%%%%%%%%%%%%%%%%%%%%%%%%%%%%%%%%%%%%%%%%%%%%%%%%%%%%%%%%%%%%%%%%%%%%%%%%%

\end{document}